# Wireless Powered Communication Networks: TDD or FDD?

Arman Ahmadian, Hyuncheol Park

School of Electrical and Electronics Engineering, KAIST

{a.ahmadian, hcpark}@kaist.ac.kr

## Abstract

In this paper, we compare two common modes of duplexing in wireless powered communication networks (WPCN); namely TDD and FDD. So far, TDD has been the most widely used duplexing technique due to its simplicity. Yet, TDD does not allow the energy transmitter to function continuously, which means to deliver the same amount of energy as that in FDD, the transmitter has to have a higher maximum transmit power. On the other hand, when regulations for power spectral density limits are not restrictive, using FDD may lead to higher throughput than that of TDD by allocating less bandwidth to energy and therefore leaving more bandwidth for data. Hence, the best duplexing technique to choose for a specific problem needs careful examination and evaluation.

## 1. Introduction

Wireless energy transfer (WET) has recently drawn significant interest as an enabling technology for the next generation of wireless devices. We consider a hybrid access point (HAP) that delivers wireless energy to a user, which harvests the energy and consumes it to transmit some information back to the HAP.

Optimization of throughput in such a network has been considered in the literature. In [1] an FDD WPCN system and in [2] a TDD WPCN system has been considered. Reference [3] considers throughput optimization in a TDD single antenna multiple user system. Nevertheless, choosing a duplexing scheme for a WET-enabled WPCN system involves many fundamental trade-offs which, to the authors' knowledge, have not yet been fully considered. Some limitations giving rise to these trade-offs include

- Due to hardware limitations, the HAP transmitter's maximum instantaneous transmit power is limited;
- To avoid interference, regulations limit the power spectral density transmitted at each frequency;
- The amount of total time-average power allocated to the WPC network can be limited.

However, depending on the operation mode of the system, some of these constraints pose more restrictions than the others. For example, while WET can utilize the whole available bandwidth in a TDD system, it cannot be enabled continuously. This imposes a strict constraint on system performance as in many cases the HAP transmitter's maximum instantaneous power is the main limitation, not the total time-average consumed power. On the other hand, in FDD systems, WET is enabled continuously. Yet, power has to be transmitted only in a portion of the total allocated bandwidth. This means that in such systems, we must consider the FCC regulations for power spectrum density more seriously.

The second section of the paper presents the common system model; i.e. part of the system model that applies to both systems. In the next section, we will analyze the TDD system and derive its optimization problem. Then the FDD scheme will be discussed and its optimization problem will be derived. Finally, we will present simulation results and the conclusion.

## 2. Common System Model

The network consists of one HAP and a single user. The HAP is connected to a wired power supply and delivers wireless power in the downlink (DL) to the user. The user, on the other hand, harvests the energy and uses it to transmit some information back to the HAP in the uplink (UL). The DL and UL channels for the user are denoted by complex random variables $\tilde{h}$ and $\tilde{g}$ respectively with channel power gains $h = |\tilde{h}|^2$ and $g = |\tilde{g}|^2$. We assume block flat fading model holds where $\tilde{h}$ and $\tilde{g}$ remain constant during one block of length $T$ but can change from one block to the other. The total bandwidth is assumed to be $w_0$. The HAP power amplifier transmits $p_d$ watts of power, which, we assume, cannot be greater than $p_{\max}$. We consider $s_{\max}$ as the maximum allowed power spectral density of the transmitted power-bearing signal due to spectral regulations.

## 3. TDD-based WPCN

In TDD, we use the "harvest-then-transmit" protocol, where the HAP first broadcasts wireless energy to the user in the DL in $\tau T$ seconds, and then the user transmits its information to the HAP in the UL in $(1 - \tau)T$ seconds using the harvested energy where $\tau \in [0\ 1]$ is the WET time ratio. In Fig. 1, the frame structure of the TDD scheme is illustrated. The energy harvested by the user in the DL is $\epsilon = \tau T p_d h$ which we assume is saved in a lossless and infinite capacitor and is used in the next wireless information transmission (WIT) phase. The transmit power of the user at WIT phase is $p_u = \epsilon/((1-\tau)T)$. Therefore, we can express the throughput of the user as

$$R(\tau) = (1-\tau)w_0 \log_2\left(1 + \gamma \frac{\tau}{1-\tau}\right) \qquad (1)$$

where $\gamma = ghp_d/\sigma^2$ and $\sigma^2$ represents the variance of the channel noise which we assume is modeled by a circularly symmetric complex Gaussian (CSCG) random variable. Our purpose is to maximize the throughput. As a result, we have the following optimization problem

$$\begin{aligned}\underset{\tau}{\text{maximize}} \quad & R(\tau)\\ \text{subject to} \quad & 0 \leq \tau \leq 1\\ & p_d = w_0 s\\ & s < s_{\max}\\ & p_d < p_{\max}\end{aligned}$$

The first constraint is because the WET phase length cannot be negative or longer than the frame length. The second constraint is because the whole allocated bandwidth is used for power transmission in TDD. The third constraint is because of the maximum allowed power spectral density. Finally, the fourth constraint is because the DL transmit power cannot be larger than the maximum allowed transmit power of the HAP power amplifier.

4. FDD-based WPCN

In FDD, the energy is transmitted and consumed simultaneously. The HAP broadcasts wireless energy to the user in a bandwidth of $\beta w_0$ and the user transmits its information in a bandwidth of $(1-\beta)w_0$ where $\beta \in [0\ 1]$ is the WET bandwidth ratio. In Fig. 1, the frame structure of the FDD scheme is illustrated. Here, we do not have the constraint of time as we did in TDD. Instead of time, the bandwidth is divided between the UL and the DL. The energy harvested by the user is: $\epsilon = T p_d h$. On the other hand, due to the reduced UL bandwidth, the channel noise variance is now $\sigma^2(1-\beta)$. Furthermore, the transmit power of the user is $p_u = \epsilon/T$. As a result, the achievable UL throughput of the user can be expressed as

$$R(\beta) = (1-\beta) w_0 \log_2\left(1 + \gamma \frac{\beta}{1-\beta}\right) \quad (2)$$

where $\gamma = ghs w_0/\sigma^2$. This equation has the same form as that for TDD. The optimization problem pertaining to FDD is as follows

$$\begin{aligned}\underset{\beta}{\text{maximize}} \quad & R(\beta)\\ \text{subject to} \quad & 0 \leq \beta \leq 1\\ & p_d = \beta w_0 s\\ & s = s_{\max}\\ & p_d < p_{\max}\end{aligned}$$

The first constraint is because the WET bandwidth cannot be negative or longer than the whole bandwidth. The second constraint is because the allocated bandwidth for power transmission is $\beta$ times the whole bandwidth $w_0$. The third constraint is because in FDD, the HAP transmits power with maximum power spectral density $s_{\max}$. Finally, the fourth constraint is because the DL power cannot be larger than the maximum allowed transmit power of the HAP power amplifier.

5. Simulation Results

We will solve both optimization problems numerically. Consider a case where $\sigma^2 = -120\text{dBm}$, $p_{\max} = 0.1\text{W}$, $g = h = 10^{-6}$, $T = 1\text{ms}$, $s_{\max} = 10^{-5}\text{WHz}^{-1}$, and $w_0 = 10\text{KHz}$. Note that here $p_{\max}$ and $w_0 s_{\max}$ are equal. Both optimization problems give similar results: $\tau = \beta = 0.27$ and $R = 38.3\text{Kbps}$. Yet, increasing the maximum power spectral density to $s_{\max} = 10^{-4}\text{WHz}^{-1}$ gives $s_{\max} w_0 =$

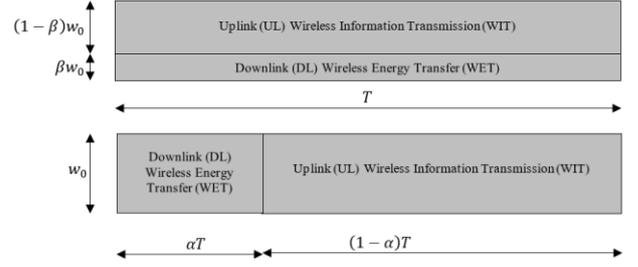

Fig. 1. WPCN duplexing frame structure. *Top*: FDD, *Bottom*: TDD

1W. This does not affect the result for TDD. However, it changes the optimal value of FDD WET bandwidth ratio to $\beta = 0.1$ and the data rate increases to $R = 61.3\text{Kbps}$. Similarly, decreasing the maximum transmit power to $p_{\max} = 0.01\text{W}$ changes WET time ratio in TDD to $\tau = 0.42$ and decreases the data rate to $R = 17.7\text{Kbps}$. However, this change in FDD leads to $\beta = 0.1$ and $R = 32.4\text{Kbps}$ which is less decrease compared to TDD.

6. Conclusion

From the discussion and the results, we can see that
- The objective function in the UL throughput maximization problem of TDD and FDD WPCN systems are essentially identical which means that throughput can be the same in some cases;
- FDD transmission scheme requires CSI feedback, which adds some complication to the system; whereas, the TDD scheme can use reciprocity and hence does not need CSI feedback;
- On the other hand, higher permissible power spectral density always increases the throughput of the FDD system, but does not necessarily does so in the TDD system;
- Similarly, decreasing the maximum HAP power decreases the throughput of both systems. Nevertheless, the decrease in the FDD system is less than or equal to that in the TDD system.

7. Acknowledgement

This work was supported by the National Research Foundation of Korea (NRF) grant funded by the Korea government (MSIT) (2017R1A2B4009853).